\documentclass{aastex631}

\usepackage{graphicx}	% Including figure files
\usepackage{amsmath}	% Advanced maths commands
\usepackage{commath}

\newcommand{\hypgeo}[2]{%
  \operatorname{%
    {\vphantom{\mathnormal{F}}}_{#1}%
    \kern-\scriptspace
    \mathnormal{F}_{#2}%
  }%
}

\begin{document}

\title{Black hole encircled by a thin disk: fully relativistic solution\footnote{Dedicated to our teacher, colleague and friend, Professor Oldřich Semerák, on the occasion of his 60th orbit around the Sun. Oldřich's well-known lectures -- famous not only among us (former) students of `math-phys' in Prague -- were the first impulse that led us to devote our studies to general relativity.}}

\author[0000-0002-9228-0788]{Petr Kotlařík}
\affiliation{Institute of Theoretical Physics, Faculty of Mathematics and Physics, Charles University,  Prague, Czech Republic}

\author[0000-0002-0278-7009]{David Kofroň}
\affiliation{Institute of Theoretical Physics, Faculty of Mathematics and Physics, Charles University,  Prague, Czech Republic}



\begin{abstract}

We give a full metric describing the gravitational field of a static and axisymmetric thin disk without radial pressure encircling a Schwarzschild black hole. The disk density profiles are astrophysically realistic, stretching from the horizon to radial infinity, yet falling off quickly at both these locations. The metric functions are expressed as {\em finite} series of Legendre polynomials. Main advantages of the solution are that (i) the disks have no edges, so their fields are everywhere regular (outside the horizon), and that (ii) {\em all} non-trivial metric functions are provided analytically and in closed forms. We examine and illustrate basic properties of the black-hole—disk space-times.

\end{abstract}



\section{Introduction} \label{sec:intro}

Black holes are once again in the spotlight of astrophysical interest thanks to the unprecedented improvement of imaging techniques in the last couple of years. The key aspect in imaging a black hole shadow is the presence of accreting matter. Having nonvanishing angular momentum, such matter usually forms an accretion disk inspiraling into the black hole. The usual simplification of this model is neglecting the disk self-gravitation because the real accretion disks are supposed to be much lighter than the central black hole. However, in many situations, this is not necessarily true, and many disk properties such as stability and positions of significant orbits are sensible to the gravitational field. Thus, we should include also the disk's own gravity in our considerations.

Nevertheless, the complexity of Einstein's equations makes the treatment in the most general case almost impossible using just analytical methods. We have to simplify the problem at some point. A great source of difficulties is the presence of overall net rotation and the resulting dragging effects. But if the rotation can be neglected or if it is compensated by e.g. two equal counter-orbiting streams, the situation is much simpler. Namely, any \textit{static} and \textit{axially symmetric} vacuum spacetime (or any non-empty spacetime where the condition $T^\rho_\rho + T^z_z=0$ for the stress-energy tensor holds) can always be described by a Weyl-type metric
\begin{equation}
	\dif s^2 = - e^{2\nu} \dif t^2 + \rho^2 e^{-2\nu} \dif\phi^2 + e^{2\lambda - 2\nu} (\dif\rho^2 + \dif z^2) \;,
\end{equation}
where $t, \rho, \phi,  z$ are Weyl coordinates and functions $\nu, \lambda$ depend solely on $\rho, z$. Outside sources, the function $\nu$ satisfies 3 dimensional Laplace equation (therefore $\nu$ is the counterpart of Newtonian gravitational potential), and for $\lambda$ we have
\begin{equation}
	\lambda_{,\rho} = \rho ( \nu_{,\rho}^2 - \nu_{,z}^2) \;, \qquad \lambda_{,z} = 2\rho \nu_{,\rho} \nu_{,z} \;.
	\label{eq:lambdaLineIntegral}
\end{equation}
Just like in electrostatics and Newtonian gravity, in order to find the external gravitational field in GR, we have to solve the Laplace (or Poisson) equation. But that is only the (Newtonian) part of the whole story. The field is not fully determined only by the potential -- there is also the second metric function $\lambda$ which influences the geometry in the meridional $(\rho, z)$ plane. This feature can significantly deviate from the Newtonian picture. Moreover, while the linearity of the Laplace equation means a considerate simplification in obtaining the field generated by more than one individual source (the potentials simply add up), the non-linear nature of Einstein's equations exhibits itself in the second metric function $\lambda$.

The first static superposition of a black hole and a thin disk (inverted solution of \cite{morgan_morgan_1969}) was considered by \cite{lemos_letelier_1993} and further studied in e.g. \cite{semerak_2003}. Later, \cite{semerak_2004} found a potential of a disk with a general power-law density profile in terms of infinite series and studied its properties when superposed with a black hole. In a recent revision of this topic in \cite{kotlarik_2022}, we provided this result in closed-form. All these disks are thin, infinite, and have an inner rim, where higher derivatives of curvature are singular. Also, only the Newtonian part (i.e. gravitational potential) of the superposition is given analytically; the second metric function $\lambda$, if needed, has to be solved numerically using (\ref{eq:lambdaLineIntegral}). Here, we provide exact full relativistic solutions describing superposition of a black hole with a disk which extends from the horizon to infinity. The fields are everywhere regular, and both metric functions of the superposition are obtained in closed-forms. It may seem inappropriate to consider disks which reach down to the horizon, since the stationary horizons cannot host any matter, and since quasi-stationary accretion disks are assumed to end somewhere about the innermost stable circular orbit. However, (i) our disk densities drop to zero towards the black hole, so there is really no matter at the very horizon, and (ii) the accretion disks around astrophysical black holes indeed continue towards the horizon (at least partially), though their matter is infalling there rather than orbiting in a quasi-circular regime. (Below the last stable orbit, the disk should not be interpreted in terms of stationary circular motion.) Hence, it is in fact more realistic to model the gravitational field of an accreting black hole employing a disk that does have a certain modest density going down to the horizon.

Before we proceed further, let us also mention other \textit{exact static} black-hole--disk superposition, where both non-trivial metric functions have been obtained. \cite{gonzalez_2009} found a family of infinite disks with an inner edge suitable for the superposition with a black hole. However, \cite{gleiser_2012} showed that the disks have problematic physical interpretation. More realistic result was obtained by \cite{vieira2020} using `displace, cut and reflect' method applied to solutions of $N$ black holes (singular rods in Weyl coordinates) arranged in a linear chain on the symmetry axis.

The paper is organized as follows. First, in the Section \ref{sec:KTdisks}, we review a family of thin infinite (in extent) -- but of finite mass -- disks due to \cite{kuzmin_1956} and \cite{toomre_1963}, which were later studied in the context of GR by \cite{bicak_1993}. Then, by inversion (Kelvin transformation) of the potential with respect to some Weyl radius, we find the potential describing, again, thin infinite disks of finite mass, but in this case, their surface densities are zero at $\rho = 0$. The same potential has been found before, as a special case, by \cite{vogt_letelier_2009} via superposition within the Kuzmin-Toomre family. Here, we contribute to this result with the second metric function $\lambda$ and thus obtain full relativistic solutions describing inverted Kuzmin-Toomre disks. Moreover, in Section \ref{sec:VLdisks}, we also provide exact solutions of $\lambda$ to all `single rings' found in \cite{vogt_letelier_2009} -- disks similar to the inverted Kuzmin-Toomre ones, but their surface density falls faster. All these disks are suitable for the superposition with a central black hole, thanks to their `annular' character. This is done in Sec. \ref{sec:superpositionwithBH} where, again, both metric functions of the black-hole--disk superposition are derived. In the subsequent Sections \ref{sec:horizonembedding} and \ref{sec:physicalproperties} we study some properties of the resulting gravitational field, namely the disk's influence on the black-hole horizon, radial profiles of densities, pressures and circular speeds in the equatorial plane. Finally, in Section \ref{sec:conclusions} we make several concluding remarks.

\section{Kuzmin-Toomre disks and their inversion}
\label{sec:KTdisks}

The $n$-order Kuzmin-Toomre disk, $n \geq 0$, is described by the potential \citep{kuzmin_1956,toomre_1963,evans_deZeeuw_1992,bicak_1993} 
\begin{equation}
	\nu_{n} = -\frac{\mathcal{M}}{(2n - 1)!!} \sum_{k=0}^n \frac{(2n-k)!}{2^{n-k}(n-k)!} \frac{b^k}{r_b^{k+1}} P_k\left( |\cos\theta_b| \right) \;,
	\label{eq:KTpotential}
\end{equation}
where we have denoted
\begin{equation}
	r_b^2 := \rho^2 + (|z| + b)^2 \;, \qquad |\cos\theta_b| := \frac{|z| + b}{r_b} \;,
\end{equation} 
$\mathcal{M}$ stands for the total mass of the disk, and $P_k$ are Legendre polynomials. The corresponding Newtonian density\footnote{$\nu$ is the solution of $\Delta \nu (\rho, z) = 4\pi w(\rho) \delta(z)$, where $\Delta$ is standard Laplace operator in cylindrical coordinates, $\delta$ is delta distribution.} reads
\begin{equation}
	w_{n} = \frac{(2n+1)b^{2n + 1}}{2\pi} \frac{\mathcal{M}}{(\rho^2 + b^2)^{n + 3/2}}  \;.
\end{equation}
Such a potential can be constructed by considering some mass distribution along the negative half of the axis $z<0$, then cutting the solution along the equatorial plane $z=0$ and reflecting its upper part $z>0$ to the negative part $z<0$ of the axis -- the so-called `displace, cut and reflect' method \citep{kuzmin_1956, vieira2020}. The resulting field is then symmetric with respect to the equatorial plane $z=0$ and given by
\begin{equation}
	\nu = - \int_{-\infty}^\infty \frac{W(b^\prime) \dif b^\prime}{\sqrt{\rho^2 + (|z| + b^\prime)^2)}} \;.
\end{equation}
For Kuzmin-Toomre solution, the weight function $W$ is a distribution given by some combination of the derivatives of delta distribution \citep{bicak_1993}. Note that in paper \cite{bicak_1993}, the weight function given in (2.23) does not in fact correspond to the Kuzmin-Toomre disks, but rather to the solution obtained by inversion with respect to $\rho=b$, only with different normalization -- see the explicit treatment below.
\begin{figure}
	\centering
	\includegraphics[width=\textwidth]{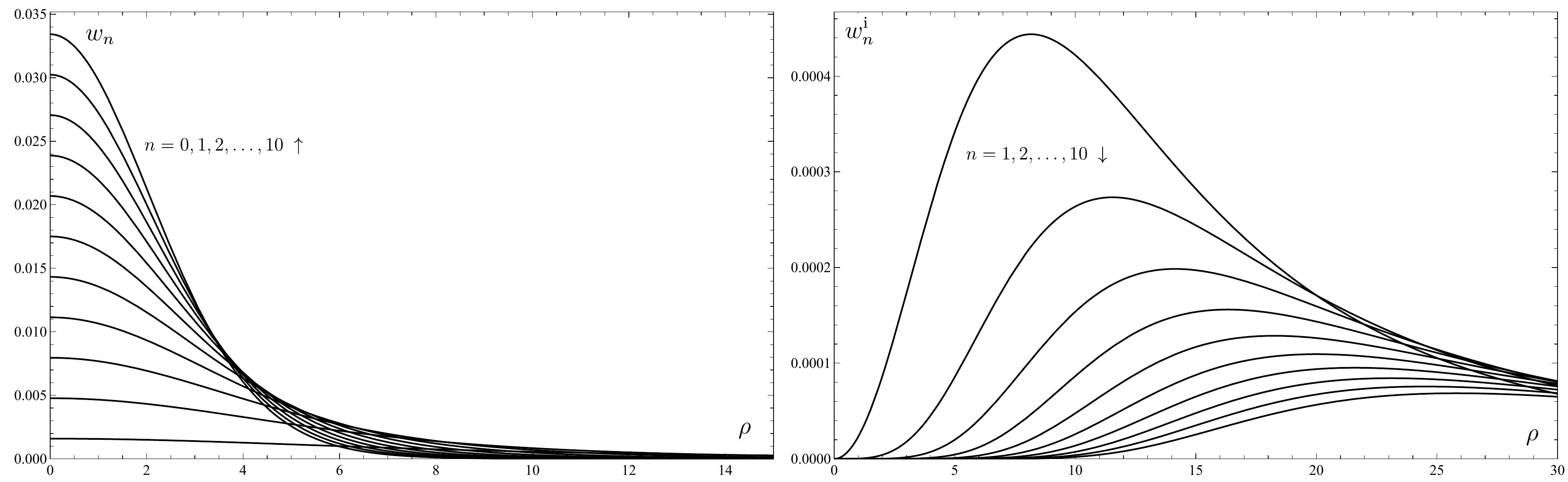}
	\caption{Radial (Newtonian) surface density profiles of the $n$th order Kuzmin-Toomre disks (\textit{left plot}) and their inversions (\textit{right plot}) corresponding to $b=10\mathcal{M}$ in both plots. The horizontal axes are in units of the disk masses $\mathcal{M}$, the vertical axes are in units of $\mathcal{M}^{-2}$.  In the \textit{left plot} the maxima lie at $\rho_\text{max} = \sqrt{2n/3} b$.}
	\label{fig:newtonianDensitiesPlotsKT}
\end{figure}

Inverted Kuzmin-Toomre disks are obtained by performing an inversion with respect to\footnote{Performing the inversion around an arbitrary radius $a$ leads to the same density profile just with different Weyl distance $b \longrightarrow a^2/b$.} $b$ (Kelvin transformation), i.e. the coordinate transformation
\begin{equation}
	\rho \longrightarrow \frac{b^2 \rho}{\rho^2 + z^2} \;, \qquad z \longrightarrow \frac{b^2 z}{\rho^2 + z^2} \;,
\end{equation} 
and the respective potential and density transformation
\begin{equation}
	\nu_{n} (\rho, z) \longrightarrow \frac{b}{\sqrt{\rho^2 + z^2}} \nu_{n} \left( \frac{b^2 \rho}{\rho^2 + z^2}, \frac{b^2 z}{\rho^2 + z^2} \right) \quad \Longrightarrow \quad w_{n} (\rho) \longrightarrow  \frac{b^3}{\rho^3}w_{n}(b^2/\rho) \;.
\end{equation}
\begin{figure}[ht]
	\centering
	\includegraphics[width=\textwidth]{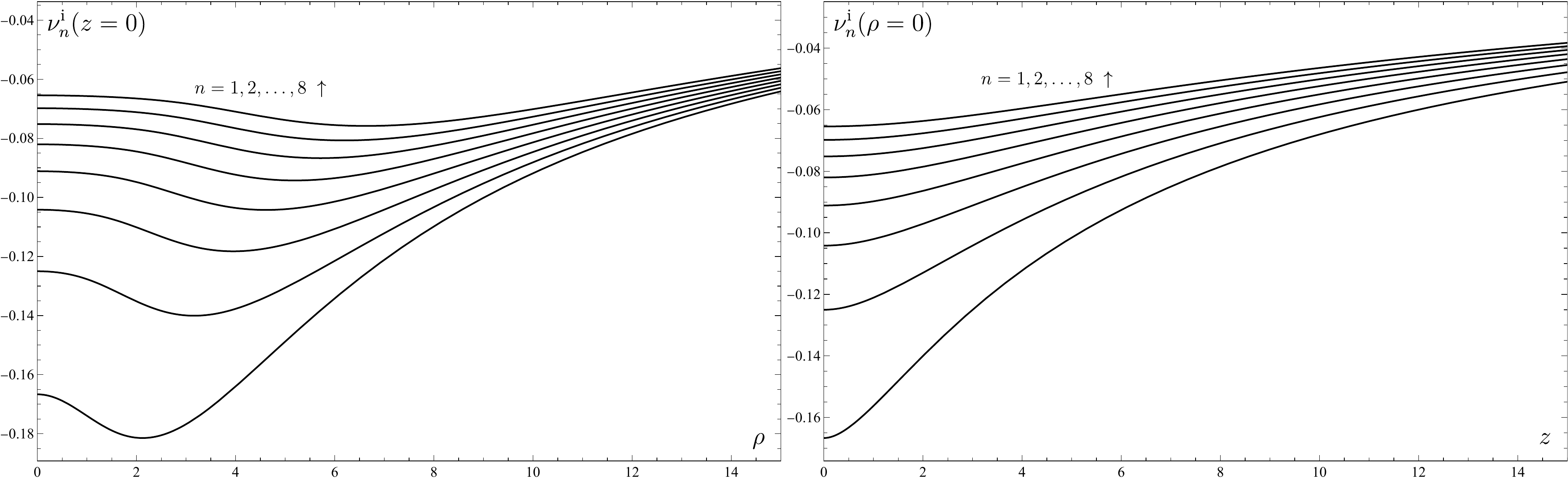}
	\caption{Coordinate plots of the inverted $n$th order Kuzmin-Toomre disk potentials in the equatorial plane (\textit{left plot}) and on the axis (\textit{right plot}) corresponding to $b = 3\mathcal{M}$. The horizontal axes are in units of the disk mass $\mathcal{M}$ while the vertical axes are dimensionless.}
	\label{fig:nuextequaaxisKT}
\end{figure}
The inversion of (\ref{eq:KTpotential}), after a suitable rearrangement of the sum, is again separable in $r_b$ and $\cos\theta_b$ and can be easily derived from (\ref{eq:KTpotential}) by (i) replacing $b \rightarrow -b$, (ii) multiplying each term by the hypergeometric function $\hypgeo{2}{1} (1+k, k-n; k-2n; 2)$, (iii) fixing the normalization, so the total mass of the disk is still $\mathcal{M}$. The inverted potential reads\footnote{Note that $$\frac{(2n-k)!}{2^{n-k}(n-k)!} \hypgeo{2}{1}(1+k, k-n; k-2n; 2) = \sum_{j=k}^n \binom{j}{k} \frac{(2n - j)!}{2^{n-j}(n-j)!}\;.$$}
\begin{equation}
	\nu_{n}^\text{i} = - \binom{n+1/2	}{n} \frac{\mathcal{M}}{(1+2n)!!} \sum_{k=0}^n \frac{(2n-k)!}{2^{n-k}(n-k)!} \hypgeo{2}{1}(1+k, k-n; k-2n; 2)   \frac{(-b)^k}{r_b^{k+1}} P_k \left( |\cos\theta_b| \right) \;,
\end{equation}
and the corresponding Newtonian density profile
\begin{equation}
	w_{n}^\text{i} = \binom{n+1/2}{n} \frac{\mathcal{M}b}{2\pi} \frac{\rho^{2n}}{(\rho^2 + b^2)^{n + 3/2}} \;.
	\label{eq:invertedKTdensity}
\end{equation}
Similarly as the original Kuzmin-Toomre disks (\ref{eq:KTpotential}), the $n$th-order potential follows also from the recurrence relation
\begin{equation}
	\nu_{n + 1}^\text{i} = \nu_{n}^\text{i} + \frac{b}{2(n + 1)} \dpd{}{b} \nu_{n}^\text{i} \;, \qquad \nu^\text{i}_0 = -\frac{\mathcal{M}}{r_b} \;.
	\label{eq:invertedKTreccurenceRelation}
\end{equation}
The shape of the disk potential along the equatorial plane and along the axis is illustrated in Fig. \ref{fig:nuextequaaxisKT}. The Newtonian surface densities are depicted in Fig. \ref{fig:newtonianDensitiesPlotsKT}.

The second metric function $\lambda_{n}$ can be also obtained following the same steps as \cite{bicak_1993}. They consist in rewriting the derivatives in Eq. (\ref{eq:lambdaLineIntegral}) in terms of $(r_b, \theta_b)$ and integrating from $r_b$ to the infinity requiring $\lambda \rightarrow 0$ when $r_b \rightarrow \infty$. Then
\begin{equation}
	\lambda_{n}^\text{i} = - \binom{n + 1/2}{n}^2 \frac{\mathcal{M}^2 \sin^2\theta_b}{\left[(1+2n)!! \right]^2} \sum_{k, l = 0}^n \mathcal{B}_{k,l} \frac{(-b)^{k + l}}{r_b^{k + l + 2}} \mathcal{P}_{k,l}(\theta_b) \;,
	\label{eq:lambdainvertedKT}
\end{equation}
where
\begin{align}
	\mathcal{B}_{k,l} &\equiv \frac{(2n - k)!(2n - l)! }{2^{2n - k - l} (n-k)! (n-l)! (k + l + 2)} \hypgeo{2}{1}(1+k, k-n; k-2n; 2) \hypgeo{2}{1}(1+l, l-n; l-2n; 2) \;,\\
	\mathcal{P}_{k,l} &\equiv (k+1)(l + 1) P_k P_l + 2(k + 1) |\cos\theta_b| P_k P^\prime_l - \sin^2\theta_b P_k^\prime P_l^\prime \;,\\
	P^\prime_k &\equiv \dod{}{|\cos\theta_b|} P_k (|\cos\theta_b|) \;.
\end{align}

\section{Vogt-Letelier disks}
\label{sec:VLdisks}

\begin{figure}
	\centering
	\includegraphics[width=.5\textwidth]{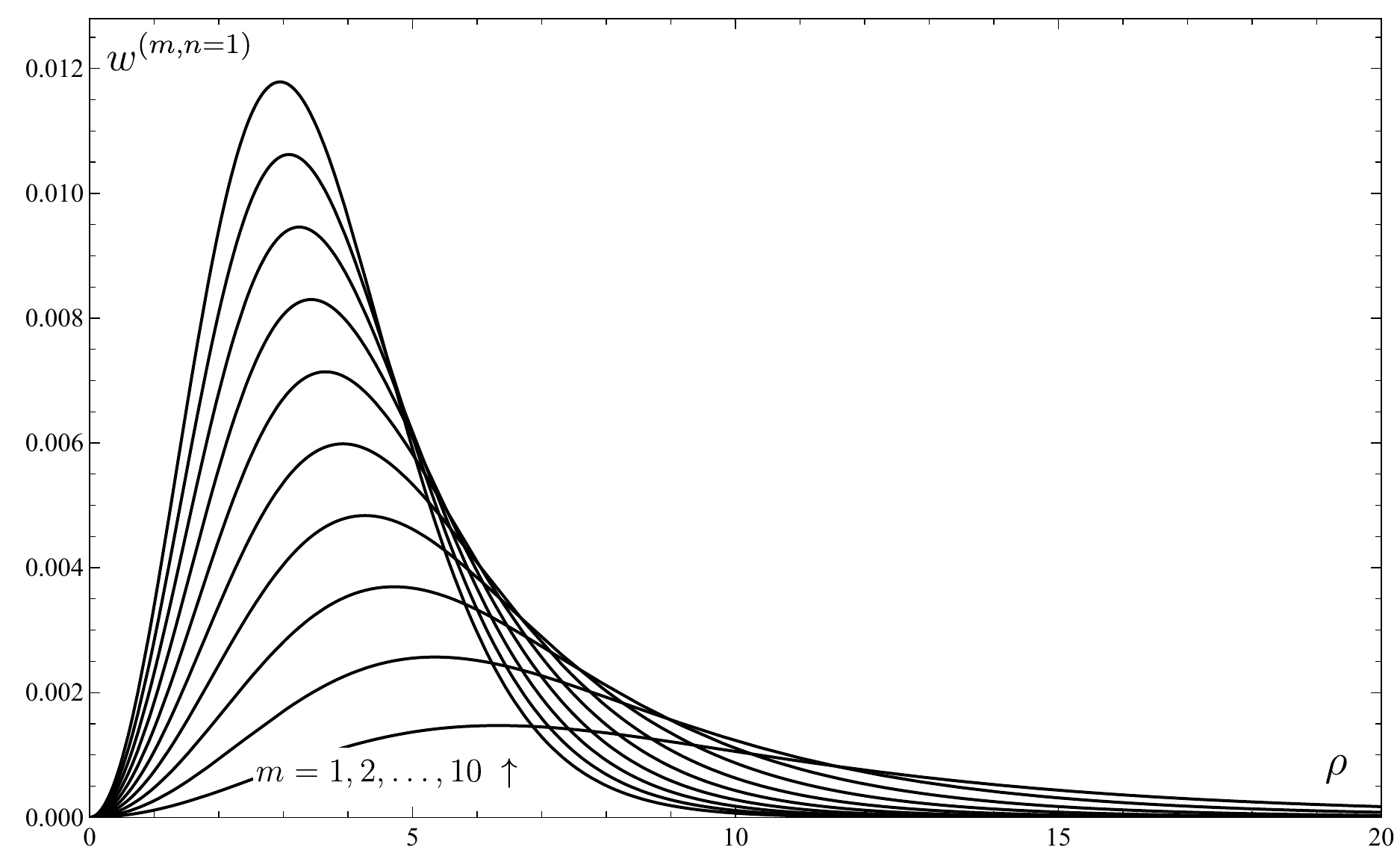}
	\caption{Newtonian surface densities of Vogt-Letelier disks for $n=1$ and $b=10\mathcal{M}$. Maxima lie at $\rho_\text{max} = 2b \sqrt{\frac{n}{m + 3/2}}$. The horizontal axis is in units of $\mathcal{M}$ while the vertical axis is in units of $\mathcal{M}^{-2}$.}
	\label{fig:newtonianDensitiesPlotsVL}
\end{figure}
\begin{figure}
	\centering
	\includegraphics[width=\textwidth]{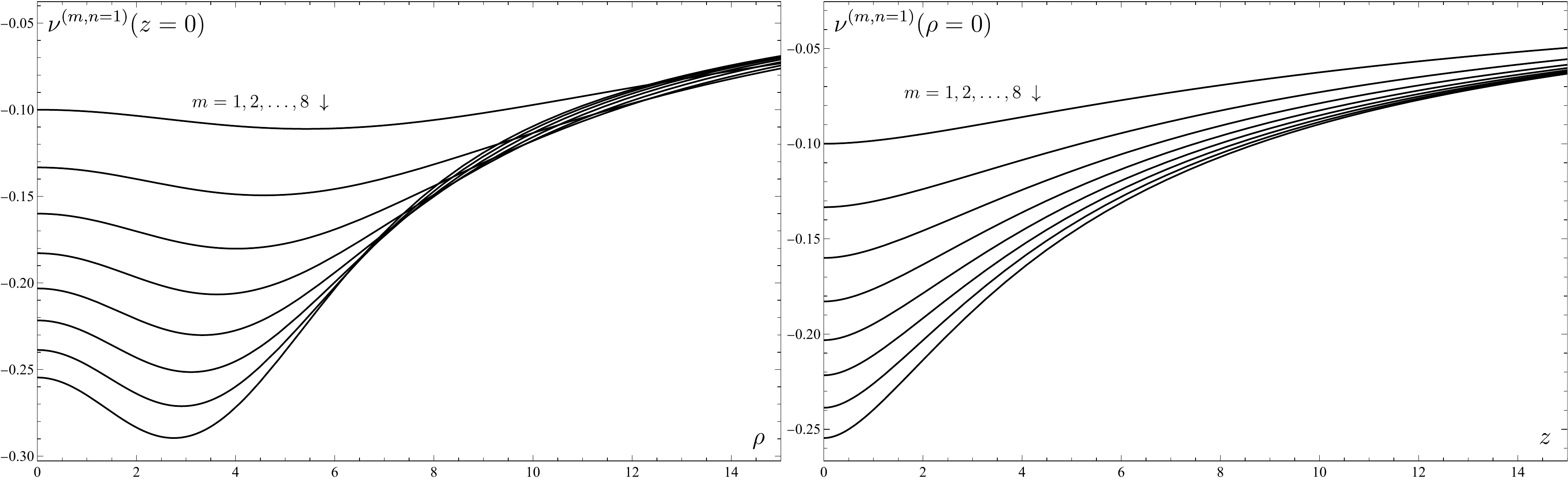}
	\caption{Coordinate plots of the Vogt-Letelier disks' potential ($n=1$) in the equatorial plane (\textit{left plot}) and on the axis (\textit{right plot}). In both plots, $b=10\mathcal{M}$, the horizontal axes are in units of $\mathcal{M}$, while the vertical axes are dimensionless.}
	\label{fig:nuextequaaxisVL}
\end{figure}

\cite{vogt_letelier_2009} obtained a broader family of disks by a superposition within the original Kuzmin-Toomre disks.  They took
\begin{equation}
	\nu^{(m, n)} =W^{(m,n)} \sum_{k=0}^n (-1)^k \binom{n}{k} \frac{\nu_{m + k}}{2m+2k+ 1} \;,
	\label{eq:nuVLoriginal}
\end{equation}
where $W^{(m, n)}$ is a normalization factor dependent only on $m, n$. Resulting disks are again of infinite extent with the surface density
\begin{equation}
	w^{(m, n)} = W^{(m, n)} \frac{\mathcal{M} b^{2m+1}}{2\pi} \frac{\rho^{2n}}{(\rho^2 + b^2)^{m + n+ 3/2}} \;,
	\label{eq:VLdensity}
\end{equation}
which behaves as $\mathcal{O}(\rho^{-2m - 3})$ at infinity, thus the total mass of each disk is again finite. We fix $W^{(m,n)}$, so the total mass of the disk is still $\mathcal{M}$, more precisely
\begin{equation}
	W^{(m,n)} =  (2m+1) \binom{m + n+ 1/2}{n} \Rightarrow 2 \pi \int_0^\infty \rho w^{(m, n)} \dif\rho  = W^{(m, n)}  \frac{\mathcal{M} b^{2m+1}}{2\pi} \int_0^\infty \frac{\rho^{2n}}{(\rho^2 + b^2)^{m + n+ 3/2}} \dif \rho = \mathcal{M} \;.
	\label{eq:Wmn}
\end{equation}
For $m=0$, we get the inverted Kuzmin-Toomre disks, which we diskussed in the previous section. disks of higher $m$ can be obtained using a recurrence relation
\begin{equation}
	\frac{(2m + 1)(2n + 3)}{2m + 2n + 3} \nu^{(m+1, n)} =  \nu^{(m, n)} + \frac{4m(n+1)}{2m + 2n + 3} \nu^{(m, n+1)} - b \dpd{}{b} \nu^{(m, n)} \;.
	\label{eq:VLrecurrenceRelation}
\end{equation}
After substitution of the Kuzmin-Toomre potential into (\ref{eq:nuVLoriginal}) the double sum can be rearranged and performed over the $k$ index. It leads to
\begin{equation}
	\nu^{(m, n)} = - (2m + 1)\mathcal{M} \binom{m + n+ 1/2}{n} \sum_{j=0}^{m + n} \mathcal{Q}_j^{(m,n)} \frac{b^j}{r_b^{j+1}} P_j(\cos\theta_b) \;,
\end{equation}
where we have already substituted $W^{(m, n)}$ from (\ref{eq:Wmn}) and
\[   
\mathcal{Q}_j^{(m,n)} = 
	\begin{cases}
		\sum_{k=0}^n (-1)^k \binom{n}{k} \frac{2^{j-k-m} (2m + 2k -j)!}{(m + k - j)!(2m + 2k + 1)!!} = \frac{2^{j-m} (2m - j)!}{(2m + 1)!!(m-j)!} \hypgeo{3}{2} \left( \frac{2m-j+1}{2}, \frac{2m-j+2}{2}, -n; \frac{2m+3}{2}, m-j+1; 1 \right) & \text{ if } j\leq m \\
		\sum_{k=j}^{m+n} (-1)^{k-m} \binom{n}{k-m} \frac{2^{j-k} (2k - j)!}{(k-j)!(2k + 1)!!} =  \frac{(-1)^{j-m} j!}{(2j + 1)!!}\binom{n}{j-m} \hypgeo{3}{2}\left( \frac{j+1}{2}, \frac{j+2}{2}, j-m-n; \frac{2j+3}{2}, j - m + 1; 1 \right) & \text{ if } j > m \;,
	\end{cases}
\]
where $\hypgeo{3}{2}$ is the generalized hypergeometric function.

The second metric function $\lambda$ is again of the same structure (\ref{eq:lambdainvertedKT}) as the original or inverted Kuzmin-Toomre disks, but now
\begin{align}
	&\lambda^{(m, n)} = -(2m + 1)^2 \binom{m + n + 1/2}{n}^2\mathcal{M}^2 \sin^2\theta_b \sum_{k, l = 0}^{m + n} \mathcal{B}^{(m,n)}_{k,l} \frac{b^{k + l}}{r_b^{k + l + 2}} \mathcal{P}_{k,l}(\theta_b) \;,\\
	&\mathcal{B}^{(m, n)}_{k.l} = \frac{\mathcal{Q}_l^{(m,n)} \mathcal{Q}_k^{(m,n)}}{k + l + 2} \;,
\end{align}
while the polynomials $\mathcal{P}_{k,l}$ are the same as in (\ref{eq:lambdainvertedKT}). A shorter expression can be found by the off-diagonal re-summation
\begin{equation}
   \lambda^{(m, n)} = -(2m + 1)^2 \binom{m + n + 1/2}{n}^2\mathcal{M}^2 \sum_{v=0}^{2(m+n)}\sum_{u=0}^v \frac{(1+u)(1+v-u)}{v+2} \frac{b^v}{r_b^{-2-v}} \mathcal{Q}_u^{(m,n)} \mathcal{Q}_{v-u}^{(m, n)}(P_u P_{v-u} - P_{1+u}P_{1+v-u}) \;,
\end{equation}
where $P_u \equiv P_u (|\cos\theta_b|)$ are the Legendre polynomials and we set $\mathcal{Q}_u^{(m, n)}=0$ for $u>m+n$. For the inverted Kuzmin-Toomre disks (i.e. $m=0$) the coefficients are reduced $\mathcal{Q}_u^{(0, n)} = \frac{(-1)^u}{(2n + 1)!!} \frac{(2n - u)!}{2^{n-u}(n-u)!} \hypgeo{2}{1} (1 + u, u-n; u-2n; 2)$.

Newtonian density profiles of several Vogt-Letelier disks are plotted in Fig. \ref{fig:newtonianDensitiesPlotsVL} and the behavior of disks' potentials in the equatorial plane and on the axis is illustrated in Fig. \ref{fig:nuextequaaxisVL}.

\section{Superposition with a Schwarzschild black hole}
\label{sec:superpositionwithBH}
	
All Vogt-Letelier disks have a clear `annular' character – their densities are exactly zero in the centre $(\rho = 0, z=0)$. Hence, it is physically reasonable to make a static superposition of such a disk with a black hole. In Weyl coordinates, the Schwarzschild black hole of the mass $M$ is a singular rod placed on the axis $( \rho=0, |z| \leq M)$ producing gravitational field described by
\begin{align}
	\nu_\text{Schw} = \frac{1}{2} \ln \frac{d_1 + d_2 - 2M}{d_1 + d_2 + 2M} \;, \qquad \lambda_\text{Schw} = \frac{1}{2} \ln\frac{(d_1 + d_2)^2 - 4M^2}{4d_1 d_2} \;,
\end{align} 
where $d_{1,2} \equiv \sqrt{\rho^2 + (|z| \mp M)^2}$. Then, the disk starts at the black-hole horizon, but the density drops to zero sufficiently quickly there. Due to the linearity of Laplace equation, the superposition of gravitational potentials is a simple sum of the individual sources, i.e. $\nu = \nu_\text{Schw} + \nu_\text{disk}$. The non-linearity of the Einstein equations manifests itself in the second metric function $\lambda$ which does not superpose that simply. In fact, for two individual sources, a black hole and a disk, we can write $\lambda  = \lambda_\text{Schw} + \lambda_\text{disk} + \lambda_\text{int}$, where $\lambda_\text{Schw}$ and $\lambda_\text{disk}$ are contributions from the black hole and the disk alone (thus satisfying (\ref{eq:lambdaLineIntegral}) with just $\nu_\text{Schw}$ and $\nu_\text{disk}$ respectively) and $\lambda_\text{int}$ is an `interaction' term for which
\begin{align}
	\lambda_{\text{int}, \rho} &= 2 \rho (\nu_{\text{Schw}, \rho} \nu_{\text{disk}, \rho} - \nu_{\text{Schw}, z} \nu_{\text{disk}, z}) \;, \\
	\lambda_{\text{int}, z} &= 2 \rho (\nu_{\text{Schw}, \rho} \nu_{\text{disk}, z} + \nu_{\text{Schw}, z} \nu_{\text{disk}, \rho}) \;.
\end{align}
Notice that these conditions are already linear in $\nu_\text{disk}$, i.e. if the potential of the disk is a sum of two components then also $\lambda$ must add in the same way. Hence, $\lambda_\text{int}$ satisfies the same recurrence relations (\ref{eq:invertedKTreccurenceRelation}) for the inverted Kuzmin-Toomre disks and (\ref{eq:VLrecurrenceRelation}) for Vogt-Letelier disks of higher $m$, namely
\begin{align}
	&\lambda^{(0,n + 1)}_\text{int} =  \lambda^{(0,n)}_\text{int} + \frac{b}{2(n + 1)} \dpd{}{b} \lambda^{(0,n)}_\text{int} \;, \qquad \lambda^{(0,0)}_\text{int} = - \frac{\mathcal{M}}{r_b} \left(\frac{d_1}{b+M} - \frac{d_2}{b-M} \right) - \frac{ 2\mathcal{M} M}{b^2 - M^2}\;,\\
	&\frac{(2m + 1)(2n + 3)}{2m + 2n + 3} \lambda_\text{int}^{(m+1, n)} =  \lambda_\text{int}^{(m, n)} + \frac{4m(n+1)}{2m + 2n + 3} \lambda_\text{int}^{(m, n+1)} - b\dpd{}{b} \lambda_\text{int}^{(m, n)}\;.
\end{align}

\begin{figure}
	\centering
	\includegraphics[width=\textwidth]{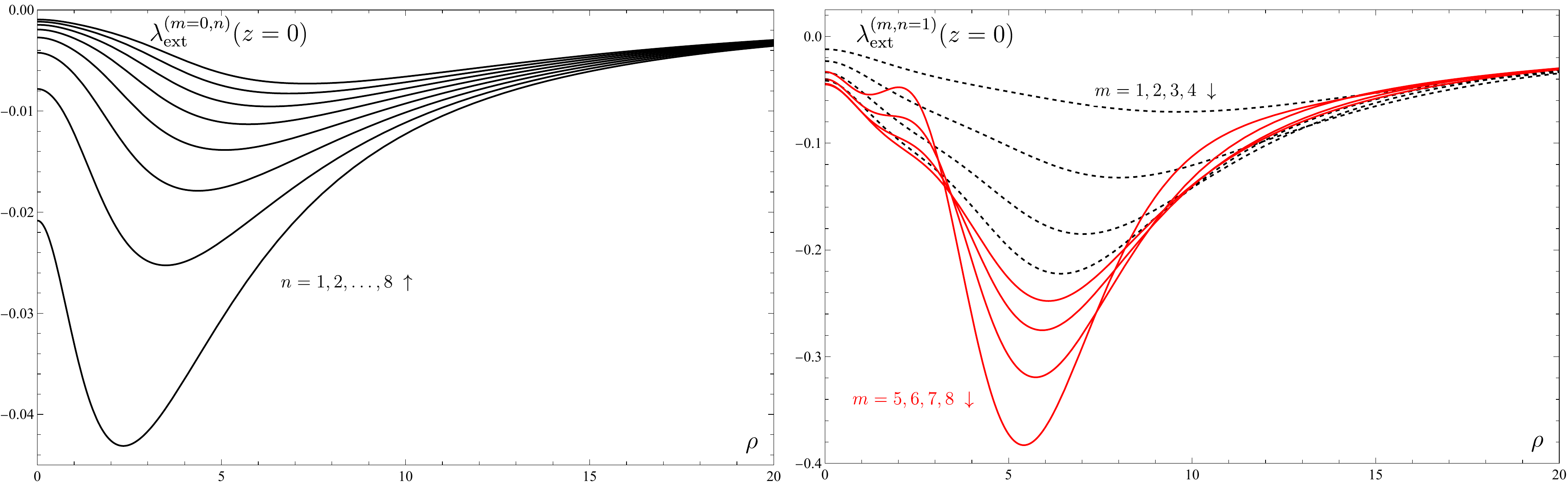}
	\caption{Equatorial plots (in Weyl radius) of $\lambda_\text{ext} = \lambda - \lambda_\text{Schw}$ of the superposition of a black hole and a disk (inverted Kuzmin-Toomre disks in the \textit{left plot}; Vogt-Letelier disks in the \textit{right plot}). Parameters of the disks are $\mathcal{M} = M$, $b=3M$ in the \textit{left plot}, and $\mathcal{M} = 3.8M$, $b=10M$ in the \textit{right plot}. The horizontal axes are in units of $M$, while the vertical axes are dimensionless. In the \textit{right plot}, the dashed black lines correspond to $m=1,2,3,4$, while the solid red lines correspond to $m=5, 6, 7, 8$. The (color) line differentiation is chosen only for the sake of clarity.}
	\label{fig:lambdaextequa}
\end{figure}

\begin{figure}
	\centering
	\includegraphics[width=.52\textwidth]{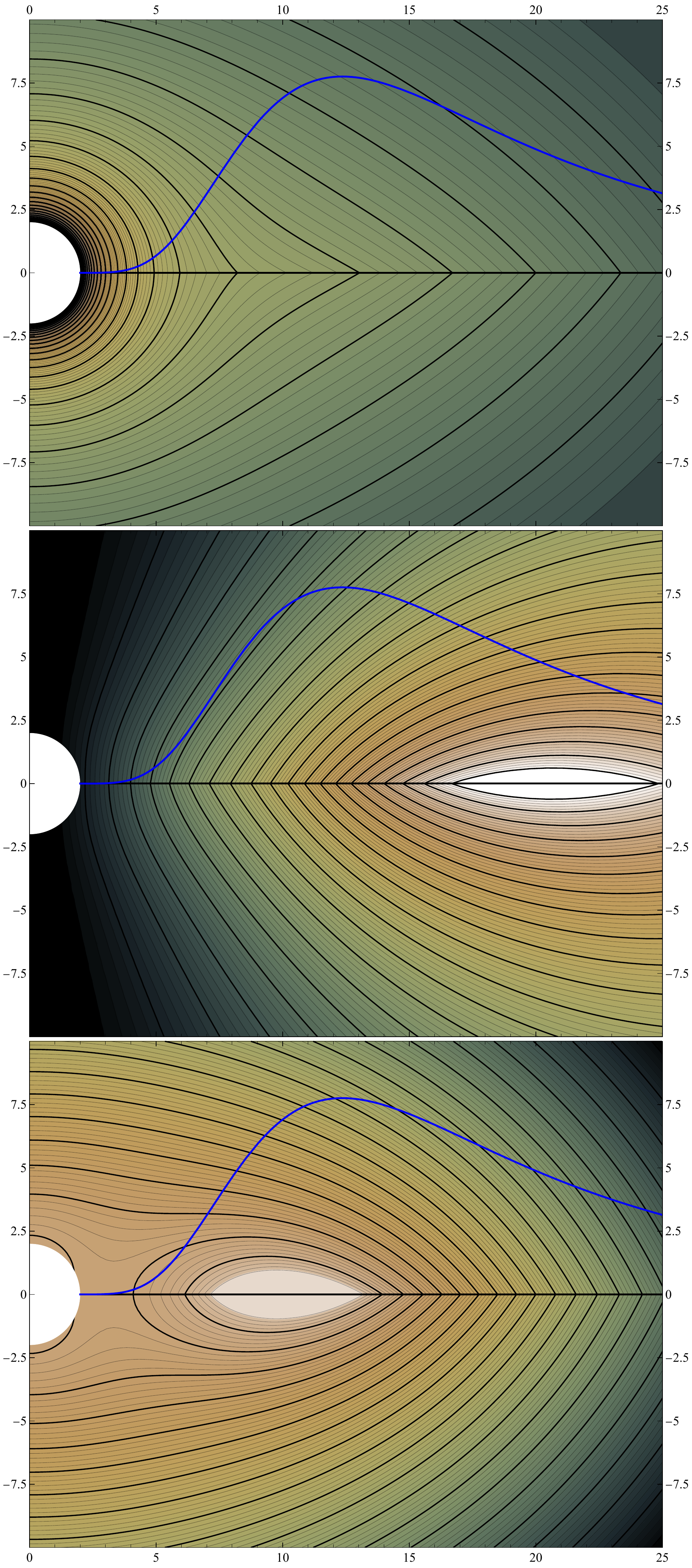}
	\caption{Meridional plane ($\phi = \text{const}$) contour plots of the superposition of the black hole and the inverted 3rd order Kuzmin-Toomre disk. Namely, the gravitational potential $\nu = \nu_\text{Schw} + \nu_\text{disk}$ (\textit{top plot}), the second metric function $\lambda_\text{ext} = \lambda - \lambda_\text{Schw}$ (\textit{middle plot}) and the difference $\lambda_\text{ext} - \nu_\text{disk}$ (\text{bottom plot}), are shown in Schwarzschild coordinates. The disk lies in the equatorial plane $z=0$ and is highlighted by a thick black line. The thick blue line shows the Newtonian density profile of the disk; the white circle in the origin shows the black hole. To illustrate how the field of the black hole is distorted by the presence of the disk, rather extreme parameters are chosen in all plots -- the mass of the disk $\mathcal{M} = 10M$ and $b=8M$.  Both axes are in units of the black hole mass $M$.}
	\label{fig:fieldcontourplots}
\end{figure}

Superposition with a black hole is best represented in Schwarzschild coordinates\footnote{Note that Schwarzschild coordinates $(r, \theta)$ are different from those coordinates with subscript $b$ used in disks' potentials.} $(r, \theta)$ introduced by
\begin{equation}
	\rho = \sqrt{r(r-2M)} \sin\theta \;, \qquad z = (r-M)\cos\theta \;.
\end{equation}
In these coordinates the black-hole horizon is rendered really spherical (at $r=2M$), the Schwarszchild contribution becomes
\begin{equation}
	\nu_\text{Schw} = \frac{1}{2} \ln\left( 1 - \frac{2M}{r} \right) \;, \qquad \lambda_\text{Schw} = \frac{1}{2} \ln \frac{r^2 - 2Mr}{r^2 - 2Mr + M^2 \sin^2\theta} \;,
\end{equation}
and the metric of the superposition reads
\begin{equation}
	\dif s^2 = - \left( 1 - \frac{2M}{r} \right) e^{2\nu_\text{disk}} \dif t^2 + e^{2\lambda_\text{ext} - 2\nu_\text{disk}} \frac{\dif r^2}{1 - \frac{2M}{r}} + r^2 e^{-2\nu_\text{disk}} (e^{2\lambda_\text{ext}} \dif\theta^2 + \sin^2 \theta \dif \phi^2) \;, \label{eq:metricInSchw}
\end{equation}
where $\lambda_\text{ext} \equiv \lambda - \lambda_\text{Schw} = \lambda_\text{disk} + \lambda_\text{int}$. The resulting field is illustrated in Fig. \ref{fig:fieldcontourplots} on several contour plots. In Fig. \ref{fig:lambdaextequa}, we show radial plots of $\lambda_\text{ext}$ in the equatorial plane.

\section{Isometric embedding of the horizon}
\label{sec:horizonembedding}

\begin{figure}
	\centering
	\includegraphics[width=.5\textwidth]{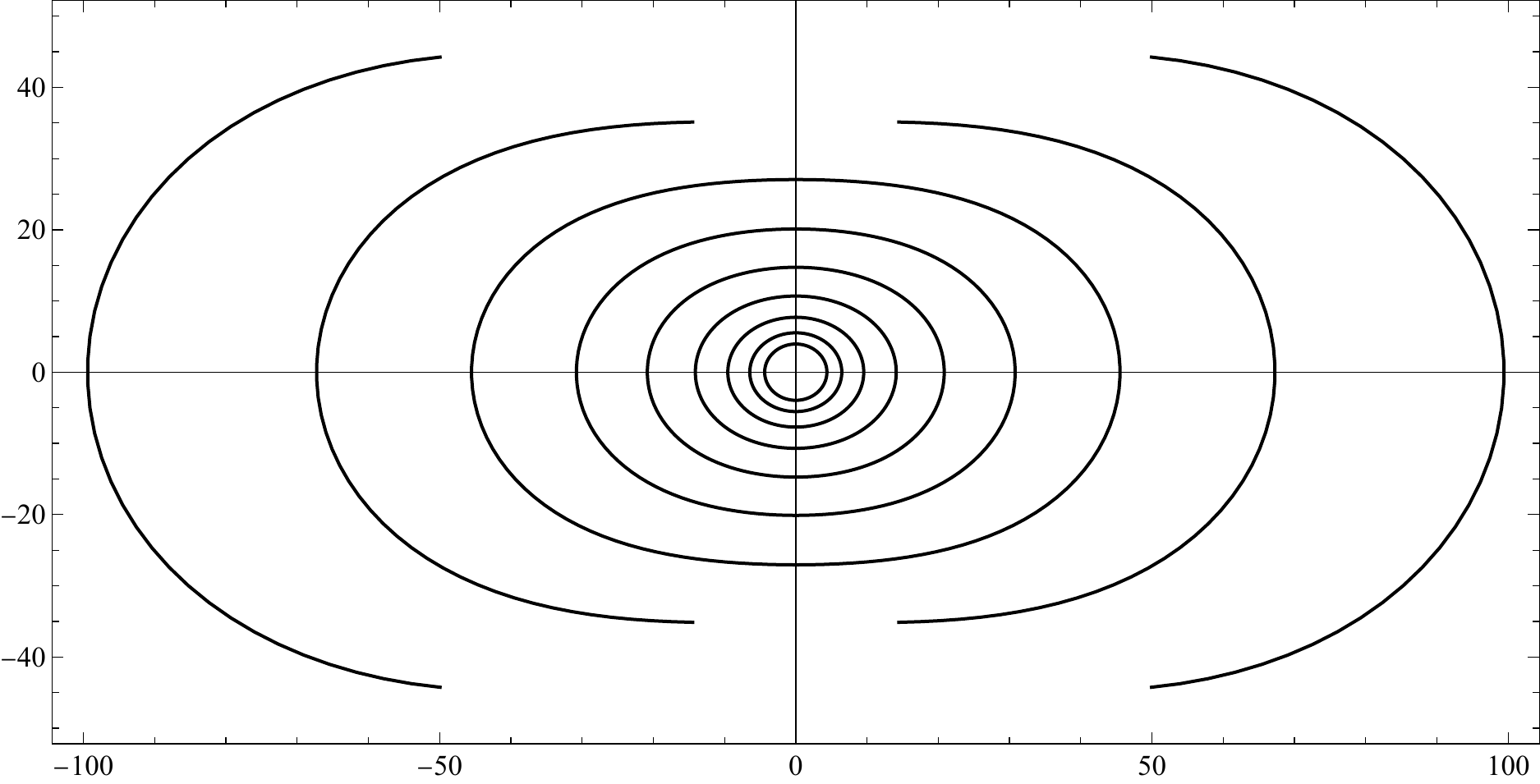}
	\caption{Meridional section ($\phi =$ const) of the isometric embedding of the black-hole horizon to Euclidean 3-space. The black hole is distorted due to the inverted 3th order Kuzmin-Toomre disk with $b=2M$. The masses of the disk range as $\mathcal{M} = 5M, 7.5M, 10M, \dots 25M$. The horizon becomes more and more flattened with increasing $\mathcal{M}$. For the last two cases, the horizon is not globally embeddable into Euclidean 3-space. Both axes are in units of $M$.}
	\label{fig:horizonEmbedding}
\end{figure}

\begin{figure}
	\centering
	\includegraphics[width=\textwidth]{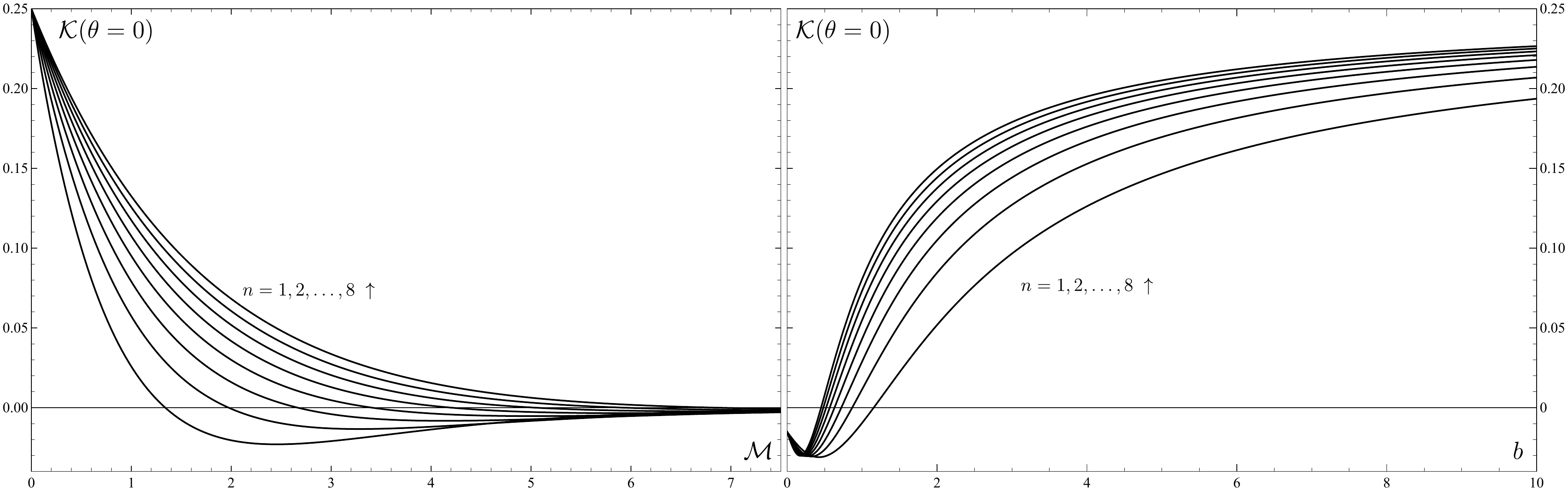}
	\caption{Gauss curvature at the horizon on the axis is plotted against the disk mass $\mathcal{M}$ (\textit{left plot}) and $b$ (\textit{right plot}). The disks are inverted Kuzmin-Toomre of the 3rd order. In the \textit{left plot} $b=0.75M$ is chosen, while in the \textit{right plot} $\mathcal{M} = 3M$ is chosen.  Gauss curvature turns negative if the disk is too massive or the density maximum is sufficiently close to the black hole. Both horizontal axes are in units of $M$ while the vertical axes are in units of $M^{-2}$.}
	\label{fig:gaussCurvature}
\end{figure}

Even though the horizon of the black hole always lies at the Schwarzschild radius $r = 2M$ and remains `spherical' in the Schwarszchild coordinates, its intrinsic geometry changes due to the presence of the surrounding matter distribution. At any given (coordinate) time, the horizon is a 2D-surface $(t=\text{konst}, r = 2M)$ with an induced metric
\begin{equation}
	\dif s^2_\text{H} = 4M^2 e^{-2\nu^\text{H}_\text{disk}(\theta)} \left[e^{4\nu^\text{H}_\text{disk}(\theta) - 4\nu^\text{H}_\text{disk}(\theta = 0)} \dif \theta^2 + \sin^2\theta \dif\phi^2  \right] \;,
	\label{eq:horizonMetric}
\end{equation}
where $\nu^\text{H}_\text{disk}$ is the limit of the disk potential at the horizon ($r\rightarrow 2M$), and where we have already used solution $\lambda^\text{H}(\theta) = 2\nu^\text{H}(\theta) - 2\nu^\text{H}(\theta = 0)$ valid on the horizon (for any \text{static} or even \text{stationary} axially symmetric spacetime). For isometric embedding to Euclidean 3-space, we use the method by \cite{smarr_1973}. It consists in rewriting the 2-metric in terms of the coordinate $\mu = \cos\theta$, i.e.
\begin{equation}
	\dif s^2 = \eta^2 \left[ f^{-1}(\mu) \dif\mu^2 + f(\mu) \dif\phi^2 \right] \;,
\end{equation}
where
\begin{equation}
	\eta = 2M e^{-\nu_\text{disk}(\mu = 1)}\;, \qquad f = e^{2\nu_\text{disk}(\mu =1) - 2\nu_\text{disk}(\mu)} (1 - \mu^2) \;.
\end{equation}
Then, the isometric embedding of the horizon 2-surface in three-dimensional Euclidean space $(x, y, z)$ is given by 
\begin{equation}
	x = \eta \sqrt{f} \cos\phi \;, \qquad y = \eta \sqrt{f} \sin\phi \;, \qquad z = \eta \int_0^\mu \sqrt{\frac{1}{f} \left( 1 - \frac{1}{4} f_{,\mu}^2 \right)} \dif \mu \;.
\end{equation}
See Fig. \ref{fig:horizonEmbedding} for the numerical results. Another useful quantity for any 2D surface is its Gauss curvature $\mathcal{K} \equiv \frac{{}^{(2)}R}{2}$, where ${}^{(2)}R$ is the corresponding 2D scalar curvature. For the metric (\ref{eq:horizonMetric}), we have
\begin{equation}
	\mathcal{K}(\theta) = \eval{\frac{1 + 3\nu_{\text{disk}, \theta} \cot\theta + \nu_{\text{disk},\theta\theta} - 2\nu_{\text{disk}, \theta}^2}{4M^2 e^{2\nu_\text{disk}(\theta) - 4 \nu_\text{disk}(0)}}}_{r=2M} \;.
\end{equation}
The Gauss curvature is plotted against $\mathcal{M}$ and $b$  in Fig. \ref{fig:gaussCurvature}. When the Gauss curvature turns negative, the horizon is not globally embeddable into Euclidean 3-space. Both, the embedding and the Gauss curvature, shows flattened horizon in the direction of the disk. 


\section{Physical properties of disks}
\label{sec:physicalproperties}

Two rather simple physical interpretations of \textit{any} static thin disks are (i) a single component ideal fluid with a certain surface density $\sigma$ and an azimuthal pressure $P$, or (ii) two identical counter-orbiting dust streams with proper surface densities $\sigma_+ = \sigma_- \equiv \frac{\sigma}{2}$ following circular geodesics.  Both characteristics follow from the diskontinuities of normal derivatives of the field over the equatorial plane
\begin{equation}
	\sigma + P = \frac{\nu_{,z}(z=0^+)}{2\pi} = w(\rho) \;, \qquad P = \frac{\lambda_{,z}(z=0^+)}{4\pi} = \frac{\nu_{,z}(z=0^+)}{2\pi} \rho \nu_{,\rho} = w(\rho) \rho \nu_{,\rho} \;,
	\label{eq:densityPressure}
\end{equation}
where $w(\rho)$ is the Newtonian density (\ref{eq:invertedKTdensity}), or (\ref{eq:VLdensity}). The second interpretation is possible only when $\sigma, P \geq 0$.

\subsection{Radial profiles of densities and azimuthal pressures}

\begin{figure}
	\centering
	\includegraphics[width=\textwidth]{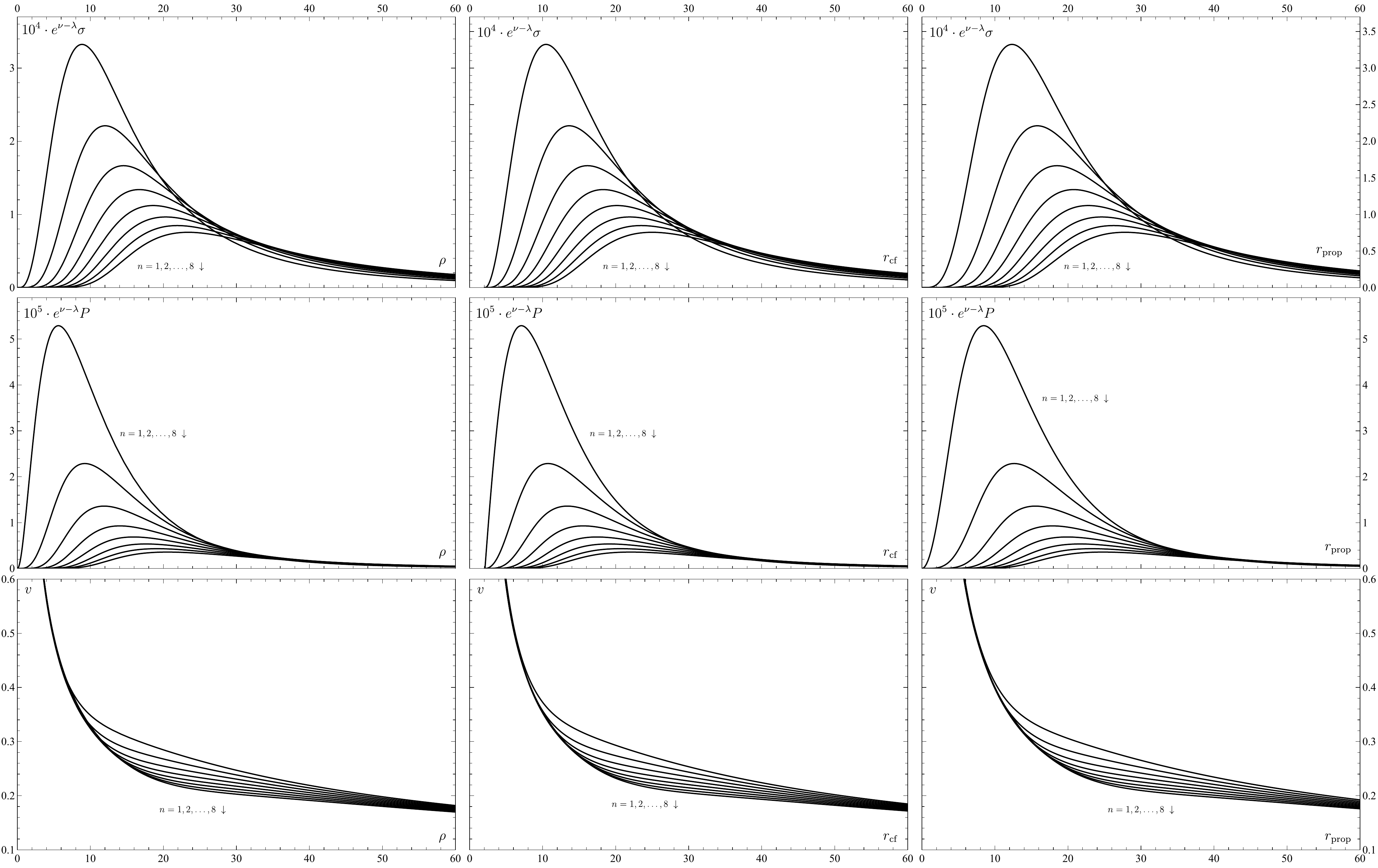}
	\caption{Profiles of disk densities (\textit{top row}), azimuthal pressures (\textit{middle row}) and circular velocities (\textit{bottom row}) in terms of the radial coordinate $\rho$ (\textit{left column}), circumferential radius $r_\text{cf}$ (\textit{middle column}) and proper distance from the horizon $r_\text{prop}$ (\textit{right column}). Eight members of the inverted Kuzmin-Toomre family ($n = 1, 2, \dots, 8$) are depicted in each plot. We chose the mass of the disks $\mathcal{M} = M$ and $b=10M$, where $M$ is the mass of the central black hole. The horizontal axes are in units of $M$ while the densities and pressures are in units $M^{-1}$ and the velocities in fractions of the speed of light.}
	\label{fig:physm0grid1}
\end{figure}

Density and pressure (\ref{eq:densityPressure}) are not, yet, quantities measured by any physical observer. Actually, an observer at rest with respect to the disk would measure the density $e^{\nu - \lambda} \sigma$ and the pressure $e^{\nu-\lambda} P$ \citep{bicak_1993}. We show radial profiles of $e^{\nu - \lambda}\sigma$ and $e^{\nu - \lambda}P$ of the inverted Kuzmin-Toomre disks in Figs. \ref{fig:physm0grid1}, \ref{fig:physm0grid2}, and of Vogt-Letelier disks ($n=1$) in Fig. \ref{fig:physn1grid1}, \ref{fig:physn1grid2}. The disks parameters in Fig. \ref{fig:physn1grid2} are rather extreme (especially for higher orders in $m$, because the mass of the disk is kept constant), but it shows the role of the multiplication factor $e^{\nu-\lambda}$, where also the second metric function $\lambda$ is present. However, the circular velocities in the disk plane (see the section below) would be superluminal in some regions, so the double-stream interpretation of such disks would not be possible.

\newpage
\subsection{Circular-velocity profiles} 

Another useful quantity is a physical speed $v$ of circular geodesics in the equatorial plane measured locally by a static observer. Such a speed is given by
\begin{equation}
	v^2 \equiv \frac{P}{\sigma} = \frac{\rho \nu_{,\rho}}{1 - \rho \nu_{,\rho}} \;,
\end{equation} 
and its time-like condition $0 \leq v^2 < 1$ covers both physical requirements for the disk -- the energy conditions and non-negativity of azimuthal pressure. In Figs. \ref{fig:physm0grid1}, \ref{fig:physm0grid2}, \ref{fig:physn1grid1}, \ref{fig:physn1vel2} we show velocity profiles for the inverted Kuzmin-Toomre disks and Vogt-Letelier disks ($n=1$).

\subsection{Coordinate and geometrical measures}

\begin{figure}
	\centering
	\includegraphics[width=\textwidth]{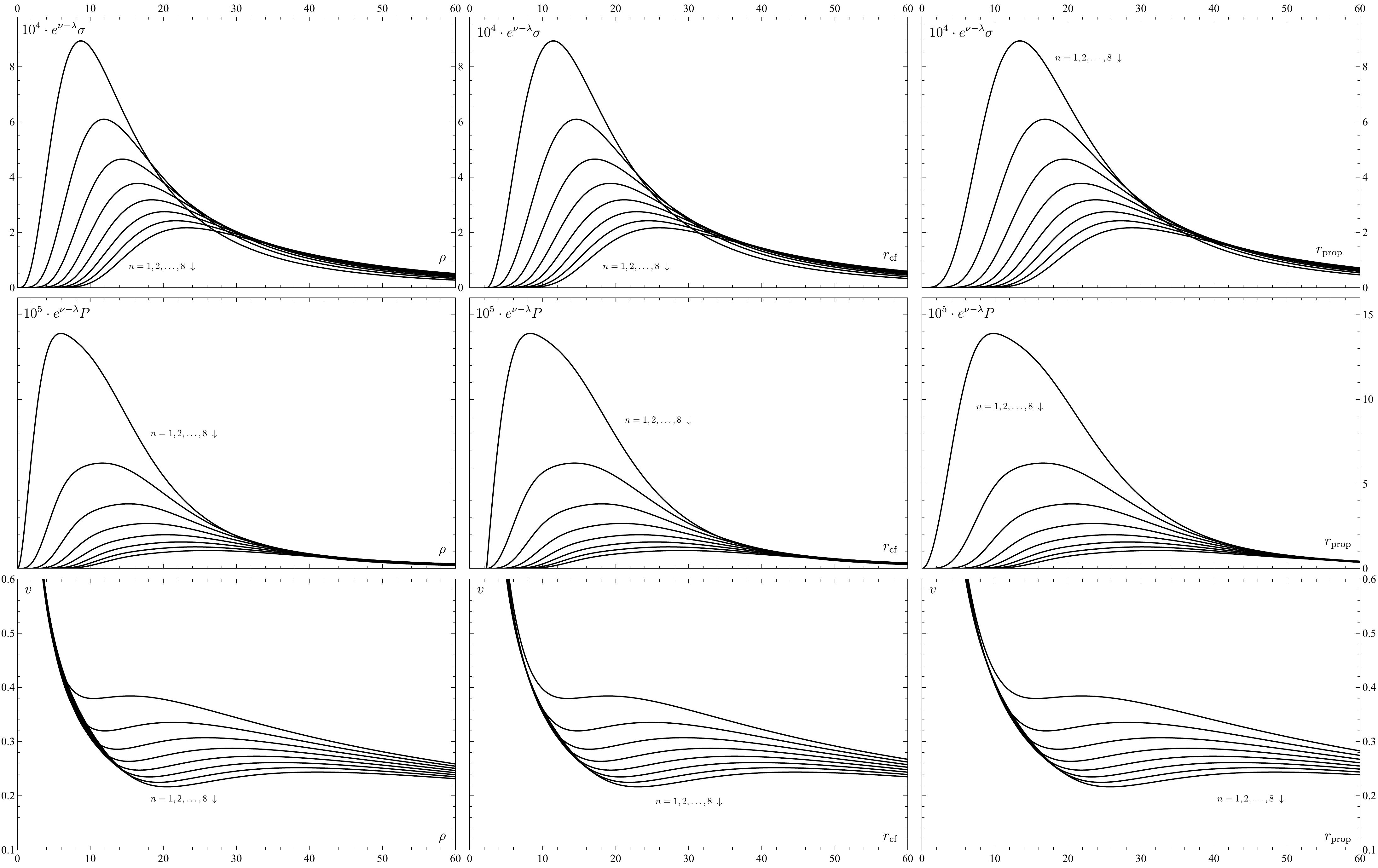}
	\caption{Same profiles as in Fig. \ref{fig:physm0grid1}. The disks belong to the inverted Kuzmin-Toomre family; the disk mass is chosen to be $\mathcal{M} = 3M$, and $b=10M$.}
	\label{fig:physm0grid2}
\end{figure}

\begin{figure}
	\centering
	\includegraphics[width=\textwidth]{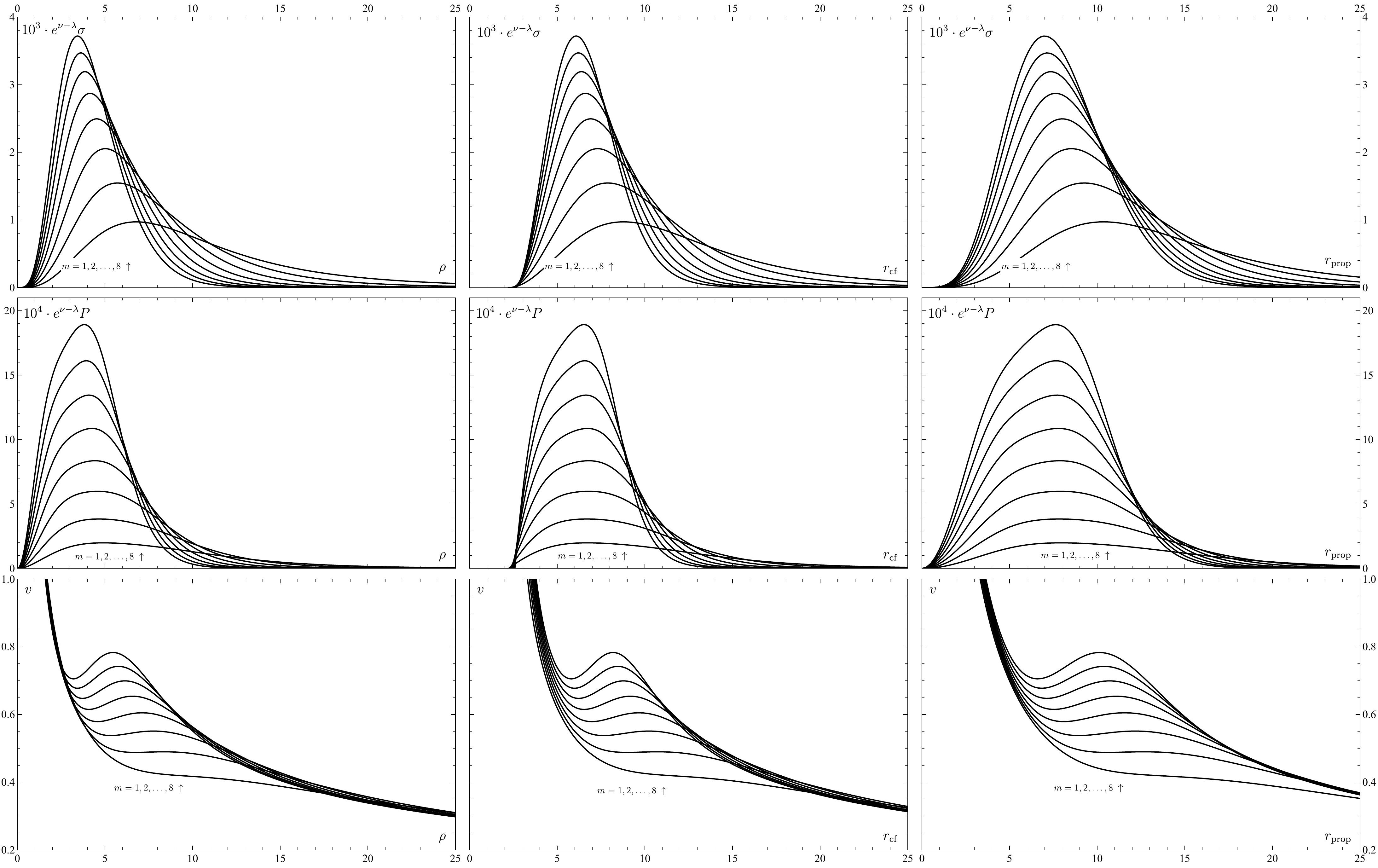}
	\caption{Same profiles as in Fig. \ref{fig:physm0grid1}, but in this figure, eight members of Vogt-Letelier disks $(m = 1, 2, \dots, 8, n=1)$ are depicted in each plot. The disks are of the same mass and $b$ as in Fig. \ref{fig:physm0grid1}, namely $\mathcal{M} = M$ and $b=10M$.}
	\label{fig:physn1grid1}
\end{figure}

\begin{figure}
	\centering
	\includegraphics[width=\textwidth]{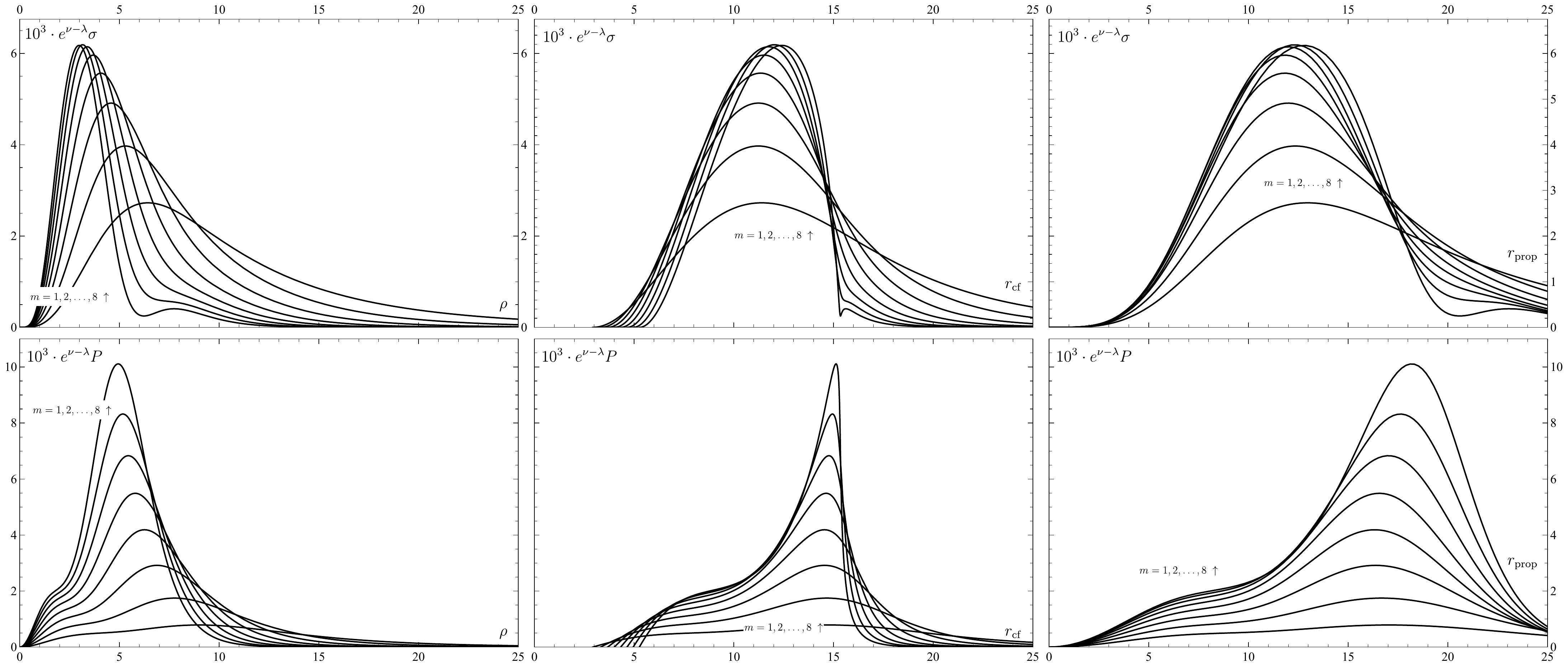}
	\caption{Profiles of disk densities (\textit{top row}) and azimuthal pressures (\textit{bottom row}) of the Vogt-Letelier disks $(m = 1, 2, \dots, 8, n=1)$ in terms of radial distances, as in Fig. \ref{fig:physm0grid1}. The rotational velocities are in separate figure, because rather extreme parameters have been chosen, namely $\mathcal{M} = 3.8M$ and $b=10M$. Meanwhile, the densities and pressures are well behaving, the rotational velocities would be superluminal in some regions of the disks, therefore the double-stream interpretation would not be possible.}
	\label{fig:physn1grid2}
\end{figure}

\begin{figure}
	\centering
	\includegraphics[width=\textwidth]{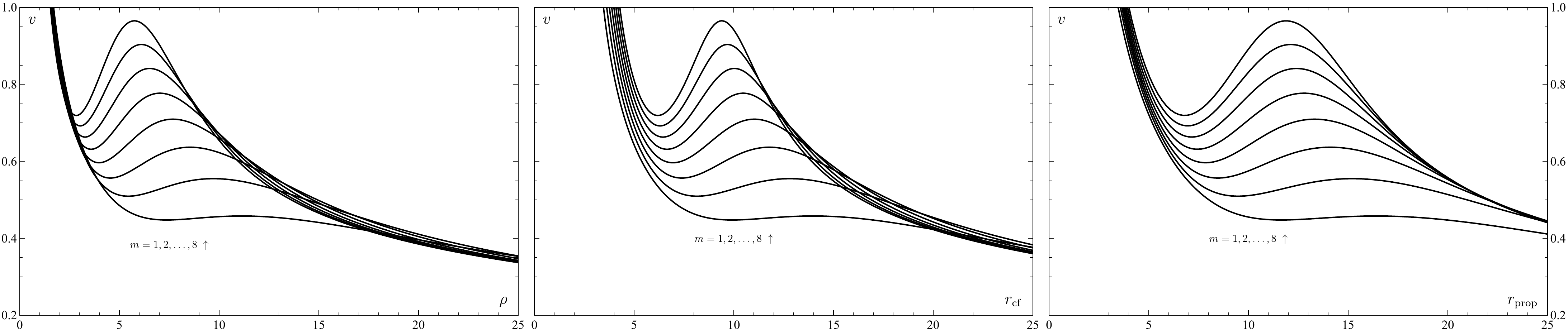}
	\caption{Profiles of circular velocities of the same disks as in Fig. \ref{fig:physn1grid2}, just with different disks mass $\mathcal{M} = 1.5M$.}
	\label{fig:physn1vel2}
\end{figure}

Most of the statements about the spacetime is given in coordinate terms. Such statements have to be taken with some caution, although the Weyl (or Schwarzschild) coordinates represents some spacetime features adequately. In particular, we should also check the physical properties of disks by employing invariant measures like circumferential radius or proper radial distance. In our case, the proper circumference corresponding to a certain $\rho$ (computed along constant $t, \rho$ and $z$) reads
\begin{equation}
	\int_0^{2\pi} \sqrt{g_{\phi\phi}} \dif \phi = 2 \pi \sqrt{g_{\phi\phi}} = 2 \pi \rho e^{-\nu} \qquad \Longrightarrow \qquad r_\text{cf} := \rho e^{-\nu} \;,
\end{equation}
where we have denoted the circumferential radius $r_\text{cf}$ in such way that the corresponding circumference is given as $2 \pi r_\text{cf}$. The proper radial distance from the black-hole horizon to a certain $\rho$, calculated in the equatorial plane $z=0$ along the remaining coordinates $(t, \phi)$ constant, is given by
\begin{equation}
	r_\text{prop} := \int_{0}^\rho \sqrt{g_{\rho\rho}} \dif\rho = \int_0^\rho e^{\lambda - \nu} \dif \rho \;.
\end{equation}
The latter integral has to be solved numerically. All quantities in Figs. \ref{fig:physm0grid1}--\ref{fig:physn1vel2} are depicted against the coordinate $\rho$ as well as both geometrical measures defined above. Note that the proper circumference is not zero at $(\rho = 0, z=0)$, instead it is the proper equatorial circumference of the black-hole horizon. As illustrated in Sec. \ref{sec:horizonembedding}, this circumference must grow due to the presence of the disk. Namely, the corresponding circumferential radius (for any black-hole--Vogt-Letelier--disk spacetime) there reads
\begin{equation}
	r_\text{cf} (\rho = 0, z=0) = \lim_{\rho\rightarrow 0} \rho e^{-\nu (\rho, z=0)} = 2 M \exp\left(\frac{(2m+1) \mathcal{M}}{b} \binom{m + n + 1/2}{n} \sum_{j=0}^{m+n} \mathcal{Q}_j^{(m, n)} \right) \;.
\end{equation}

\section{Concluding remarks}
\label{sec:conclusions}

We have found full metric describing \textit{static} and axially symmetric superposition of a black hole and a thin disk. The disks we considered have finite mass, have no radial pressure and extend from the black-hole horizon to infinity. They also have no edges, so the field is regular everywhere outside the horizon, yet the disks surface densities fall off sufficiently quickly at both extremes $\rho = 0$ and $\rho \rightarrow \infty$. In particular, we have used disks resulting from the inversion (Kelvin transformation) of Kuzmin-Toomre solutions, as well as `simple' disks due to \cite{vogt_letelier_2009} (of which the inverted Kuzmin-Toomre disks are a distinct subclass). Both metric functions are given analytically and in closed-forms. The gravitational field is illustrated in plots of the metric functions and profiles of densities, azimuthal pressures and circular velocities. The black-hole horizon was found distorted due to the presence of the disk (flattened in the direction of the disk) which is in agreement with many previous results, e.g. \cite{semerak_2004,semerak_2003,semerak_2001}, or, in the stationary case, \cite{kotlarik_2018}.

%% Also note that the akcnowlodgment environment does not support long amounts of text. If you have a lot of people and institutions to acknowledge, do not use this command. Instead, create a new \section{Acknowledgments}.
\begin{acknowledgments}
     We thank for support from GACR 21-11268S grant of the Czech Science Foundation (D.K.) and from Grant schemes at Charles University, reg.~n. CZ.02.2.69/0.0/0.0/19\_073/0016935 (P.K). We are also very grateful to Oldřich Semerák for the initial idea of inverting Kuzmin-Toomre disks and for helpful comments which improved the manuscript. 
\end{acknowledgments}


\software{Wolfram Mathematica 13}



\bibliography{bibliography}{}
\bibliographystyle{aasjournal}



\end{document}